\documentclass{kluwer}    

\usepackage{graphicx}
\usepackage{times}
\usepackage{mathptm}

\newdisplay{guess}{Conjecture}

\begin{document}                                                                                   
\begin{article}
\begin{opening}         
\title{Magnetically-Dominated Accretion Flows (MDAFs) and 
            Jet Production in the Low\/Hard State} 
\author{David L. \surname{Meier}}  
\runningauthor{David Meier}
\runningtitle{Magnetically-Dominated Accretion Flows}
\institute{Jet Propulsion Laboratory, California Institute of Technology}
\date{October 22, 2004}

\begin{abstract}
In this paper I propose that the inner part of a black hole
accretion inflow ($< 100~r_{g}$) may enter a magnetically-dominated,
magnetosphere-like phase in which the strong, well-ordered fields play a
more important role than weak, turbulent fields.  In the low/hard state
this flow is interior to the standard ADAF usually invoked to explain
the observed hot, optically thin emission.  Preliminary solutions for
these new MDAFs are presented.

Time-dependent X-ray and radio observations give considerable insight
into these processes, and a new interpretation of the X-ray power spectrum
(as arising from many disk radii) may be in order.  While an evaporative
ADAF model explains the noise power above 0.01 Hz, an inner MDAF is needed
to explain the high frequency cutoff near 1 Hz, the presence of a QPO,
and the production of a jet.  The MDAF scenario also is consistent with
the data-based, phemonenological models presented at this meeting by
several authors.
\end{abstract}

\end{opening}           

\section{Introduction:  The `Black Hole Problem'}  

A generic, robust electrodynamic model for producing most astrophysical
jets is now well understood.  It has two basic requirements:
\begin{itemize}
\item{A strong magnetic field ($V_A \equiv B/(4 \pi \rho)^{1/2} >>
C_S \equiv (\Gamma p/\rho)^{1/2}$, where $V_A$ is the Alfven speed and
$C_S$ is the sound speed) that rotates fairly rapidly ($\Omega \lesssim
\Omega_K$, where $\Omega_K$ is the Keplerian angular velocity).}
\item{Some means of loading this rotating magnetic field with plasma
at fairly high elevations ($Z \sim R$ in a cylindrical [$R,~Z,~\theta$]
coordinate system).  Early models accomplished this loading by centrifugal
action from a thin disk, but more recent studies suggest that the primary
means is thermal, (heating the plasma to roughly the virial temperature;
\opencite{Meier01}).}
\end{itemize}
The means by which this configuration produces a jet was first proposed
by \inlinecite{Bland76} and \inlinecite{Love76} and has been reviewed
recently by the author (\opencite{MKU01}; \opencite{Meier04a}).
The rotation of the magnetic field lines is retarded by the inertia
of the plasma load, creating a rotating helical field configuration.
Lorentz forces simultaneously push plasma up and out of the system along
the rotation axis and collimate the flow with magnetic hoop stress by
squeezing it toward the rotation axis, converting rotational energy
of the central engine into directed kinetic outflow energy along the
rotation axis.  The jet is accelerated to the local Alfven speed by the
rotational Alfven wave and beyond that to the local magnetosonic speed
by the toroidal magnetic pressure gradient.  The terminal velocity of
the jet is approximately equal to the escape speed at the footpoint of
the magnetic field in the rotating central engine.

\subsection{Specific Models for Stellar Jets}

While this model gives a general description of how astrophysical
jets form, it does not answer the question of how each source produces
the above two main ingredients --- the global rotating magnetic field
and the plasma loading.  For stellar sources the first requirement is
straightforward:  the global magnetic field is that produced by the
star's magnetosphere itself.  The discovery of strong magnetic fields
in pulsars and protostar systems is considered confirmation of the MHD
jet production model.

The plasma loading is more problematical.  In pulsars it has been
shown that pair production can occur in `spark gaps' in the very strong
magnetic field ($10^{11-13}$ G), producing the needed plasma inside the
magnetosphere itself (\opencite{GJ69}; \opencite{RS75}).  In protostar
systems, however, the field is not nearly as strong, so pair production
cannot operate.  Instead, the protoplanetary accretion disk is used to
pinch the field near the equator, creating an ``X-point'' \cite{S94},
where plasma is allosed to flow freely from the disk onto the rotating
stellar field.

Finally, for jets produced in collapsing supernova cores \cite{WMW02},
the plasma comes pre-loaded, since the magnetic field of the collapsing
core has become threaded into the progenitor mantle during the late
stages of stellar evolution.  It is the rapid rotation of the core that
is suddenly and rapidly generated, by the collapse of that iron core to
proto-neutron star densities.

\subsection{The Black Hole Problem}

While it is clear that black hole systems often produce fast and powerful
jets, they present a serious challenge to the electromagnetic theory of
jet production.  The plasma loading itself is a relatively easy problem to
solve.  The fact that black hole systems that produce jets are associated
with hot accretion flows \cite{Fen99} indicates that the plasma must be
loaded onto the field lines from the accretion flow by some thermal means.

However, by themselves, black holes cannot support a magnetic
field\footnote{It is true that a charged, rotating black hole has a
dipole magnetic moment \cite{MTW73}.  However, a strongly charged black
hole will induce charge separation in any surrounding plasma, accreting
charges opposite in sign to that on the hole and expelling those of like
sign.  This will effectively discharge the hole in a few light-crossing
times.}.  They can have such a field only if there is an external supply
of plasma in which currents generate magnetic flux that can then thread
the black hole.  In the absence of a significant amount of external plasma,
a black hole loses its field in a few light-crossing times \cite{TPM86}.
{\em The field, therefore, must be supported by currents in the black
hole accretion flow.} This conclusion itself presents a problem, however.
Accretion disks are believed to be weakly-magnetized plasmas in highly
turbulent, orbital flow about the black hole \cite{BH98}. How does a
global, well-ordered rotating magnetosphere develop naturally from a
turbulent accretion disk?

In addition, even if a global magnetosphere can be constructed, there
is a question as to how that magnetosphere can couple to black hole
rotation to produce a strong jet.  While the accretion disk itself can
produce rotation of the magnetic field, it cannot be the main source 
of jet power in many jet-producing supermassive black hole systems.  
Radio galaxies and quasars that
have similar optical properties, and therefore similar accretion disks,
can differ in their radio jet luminosities by factors of $10^{5-6}$.
This is most easily explained by tying jet production to rotation of
the central object, just as it is done in stellar jet-producing systems.
In addition, black hole systems are known to produce jets $\sim 30$
times stronger than those from neutron stars with similar accretion rates
\cite{Migl03}. While this comparison may be complicated by effects of the 
neutron stars' magnetic field, a strong coupling of jet production to the 
black hole spin also may be at work.  

To solve this problem we will assume here that the ``magnetic Penrose''
mechanism of extracting rotational energy is at work \cite{Koid02}: if
plasma threaded with a magnetic field enters the ergosphere, then that
plasma can be accelerated in a direction {\em opposite} to the black
hole's spin, acquiring {\em negative} energy and angular momentum in
the process.  Positive energy and angular momentum then is transferred to
the rotating magnetic field, which uses that to accelerate and collimate
the jet.

The purpose of this paper is to explore answers to the two remaining
questions: 1) how does a turbulent, magnetized disk create a global,
well-ordered magnetic field that can couple to the black hole rotation
and 2) how does the accretion disk load the field lines with plasma?

\section{Basics of Magnetically-Dominated Accretion Flows}

\subsection{What is an MDAF?}

An MDAF is an accretion flow in which the magnetic forces dominate over
the thermal and radiation forces.  In a normal accretion disk model,
the weak magnetic field creates a ``magneto-rotational instability''
(MRI) \cite{BH98} in which turbulence dominates the angular momentum
transport and the eddy turnover time $\tau_{turb}$ is shorter than the
inflow time $\tau_{inflow} \equiv R / V_R$.  A steady disk structure
develops in which magnetic field components $B_R \sim B_{\phi} \propto
R^{-5/4}$ and pressure scales as $p \propto R^{-3/2}$.  The ratio of
magnetic to thermal forces $\alpha \propto B_R ~ B_{\phi} / p$ remains
constant at $\sim 0.01 - 1.0$.

We recognize two types of magnetically-dominated accretion flows.
The first is still turbulent, but now the ratio of the time scales
is reversed: $\tau_{inflow} < \tau_{turb}$.  Small eddies continue to
transport angular momentum, but the larger ones are stretched out in the
$R$ direction before they have a chance to turn over.  In this case, $B_R
\propto R^{-5/2}$ and $B_{\phi} \propto R^{-1/2}$ decouple and $p \propto
R^{-3/2}$, so that magnetic stresses increase as $R$ decreases: $\alpha
\propto R^{-3/2}$.  We call this type of flow ``transitional'', because
it connects a turbulent flow with $\alpha < 1$ to one with $\alpha > 1$
and the MRI turned off.  If $\alpha_0$ is the value at $R_0$, and $R_1$
is the radius where $\alpha$ attains unit value, then
\begin{equation}
R_1/R_0 ~ = ~ {\alpha_0}^{-2/3}
\end{equation}
If $\alpha_0 \sim 0.3$, as is expected in advection-dominated accretion
flows \cite{NMQ98}, then $R_1/R_0 \sim 0.5$.  So, if the interior of
an ADAF becomes magnetically dominated, the transition region will be
rather narrow in radius.

In the second type of MDAF, which is a solution to ``Gammie flow''
\cite{Gam99}, MRI turbulence has ceased and the inflow is laminar along
strong magnetic field lines.  $B_R \propto R^{-3/2}$ and $B_{\phi} \propto
R^{-1}$ are still decoupled. The thermal pressure scaling depends
critically on the energy balance in the gas now, but simple models
indicate $p \propto R^{-1/2}$. So $\alpha \propto R^{-2}$ continues to
increase inward, and the flow continues to become {\em more} magnetically
dominated as it approaches the black hole. Figure 1 shows a schematic
of our low-state model and will be discussed more fully below.

\begin{figure}
\begin{center}
\centerline{\includegraphics[scale=0.34, angle=0]{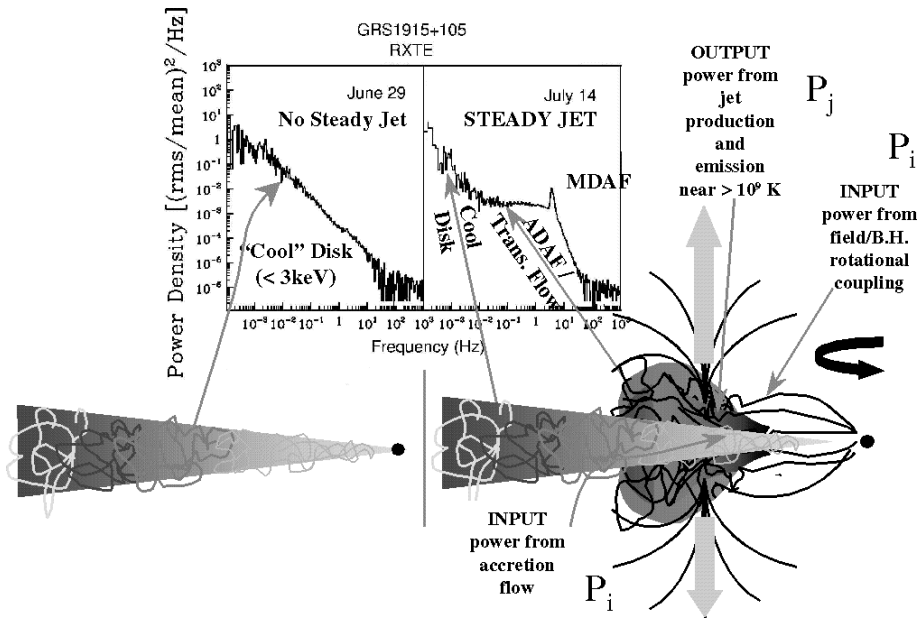}}
\caption{\label{fig:mdaf}
Schematic diagram of the MDAF model for GRS 1915+105.  Power spectrum
data are taken from Morgan {\it et al.} (1997). A cool disk fits the
power spectrum in the high state (left) when no jet is produced.  In the
hard state (right) we not only need an ADAF (extended corona) but also an
MDAF (inward-facing magnetosphere) to produce the few Hz cutoff and QPO,
and an outward-facing magnetosphere to produce the jet.  The jet will be
launched from the transition radius near $\sim 100~r_{g}$.  The input
and output powers ${\rm P}_{i,j}$ correspond to those in the Malzac {\it
et al.} model. An MDAF is also expected in some intermediate accretion
states when the cool disk is still truncated at a radius $> r_{g}$.
}
\end{center}
\end{figure}

MDAF-type solutions are seen in MRI simulations only in the plunging
region very near the black hole, not out to distances as large as $\sim
100~r_{g}$. Why? The reason may be the assumption implicit in the
simulations that the flow is radiatively inefficient.  This may be the
case for $R \gtrsim 100~r_{g}$ (where $r_{g} \equiv G M / c^2$);
in this case the temperature remains $T \lesssim 5 \times 10^9$ K.
However, inside this radius electrons radiate copiously by synchrotron,
pair production, and other relativistic processes.  While ADAF models
assume that the ion temperature can remain hot ($T_i \sim 10^{12}
{\rm K} (R/r_{g})^{-1}$), if there is strong coupling between ions
and electrons, the ions will cool to the $\gtrsim 10^9$ K temperature
as well.  It is often assumed that such cooling would lead once again
to a geometrically thin, optically thick disk.  However, these models
show that there is another solution:  cool, but still optically thin flow
along strong magnetic field lines.  It, therefore, is extremely important
to begin performing MRI simulations with a {\em real} energy equation,
including separate evolution of the ions and electrons.

\subsection{What are the Properties of MDAFs?}

The inner MDAF is an extraordinarily {\em in}efficient flow.  It is a
nearly-radial in-spiral, geometrically thick because of {\em magnetic}
pressure support \cite{Meier04a}.  Virtually all orbital angular momentum
is transferred out to $R_1$ along the strong field lines. The plasma
experiences only compressional heating and radiative cooling by electrons
and could remain quite cool;  the majority of the gravitational energy
released is converted into radial infall kinetic energy, not heat.

Because the magnetic radial channels are potentially distinct, the flow
may break up into inhomogeneous ``spokes''.  A signature of an MDAF may
be a quasi-periodic oscillation at one or more of two transition radius
frequencies: 1) the orbital/Alfven frequency
\begin{equation}
\label{eq:nuA}
\nu_A = V_A / 2 \pi R_1 = 1.1 {\rm Hz} (M_{\bullet} / 10 M_{\odot})^{-1}
\end{equation}
and 2) the MHD slow mode (``organ pipe'') frequency
\begin{equation}
\label{eq:nuS}
\nu_S = V_S / 2 \pi R_1 \approx C_S / 2 \pi R_1 = 0.02 - 0.13 ~ \nu_A
\end{equation}
or $\sim 20 - 140 {\rm mHz}$ for a $10 M_{\odot}$ black hole, depending on
the azimuthal length of the resonating magnetic tubes.  Because $\nu_S$
is excited acoustically near $R_1$ along the length of these tubes, it
may be characterized by multiple harmonics, whereas the orbital/Alfven
mode should be rather pure.

The magnetic field lines extending inward toward the black hole may
tap the hole's rotational energy if they penetrate the ergosphere.
However, in general the hole rotation rate will not match the, usually
slower, $\nu_A$.  One therefore might expect an episodic interaction,
where the field enters the ergosphere, is wound up rapidly, reconnects
in a series of rapid flares separated by the ergosphere rotation time,
and finally pulls back from the hole for a secular time.  The behavior
of SgrA at the Galactic center \cite{Gen03} is very similar to what
might be expected from a rotating black hole/MDAF interaction.

In addition to an inner magnetosphere of closed field lines reaching
toward the black hole, there also may be open field lines extending from
$R_1$ to infinity (see Fig. 1 and \opencite{Meier04a}).  The transition
radius, therefore, has all the properties necessary to launch a jet: the
base of large-scale, open, rotating magnetic field lines being loaded
with hot ADAF material. Excess angular momentum is deposited at $R_1$
by the radial field lines that connect periodically to the black hole
ergosphere.  An outflowing MHD wind/jet would be a good candidate for
carrying off this excess angular momentum.

\section{Discussion}

\subsection{MDAFs and the Low/Hard (Plateau) State of XRBs}

When the transient X-ray binary (XRB) source GRS 1915+105 is in the soft state and not
producing a jet, its photon spectrum is dominated by a cool thermal
spectrum, and its {\em power} spectrum is a rather featureless power law
of $dP/d{\nu} \propto \nu^{-4/3}$.  When the source begins to produce a
steady jet, it enters a low/hard state in which the photon emission is
dominated by a {\em non}-thermal spectrum. And the power spectrum develops
bandwidth-limited noise (a flat ($dP/d{\nu} \propto \nu^0$ shoulder with
a steep cutoff above $\sim 3$ Hz) and a QPO at 1 - 3 Hz. It is natural
to associate the non-thermal photon spectrum and bandwidth-limited
noise with the optically thin, turbulent ADAF that has formed in the
center of the disk.  But what produces the QPO, and why would the ADAF
be bandwidth-limited?  Why does its turbulence not extend all the way
to the natural frequencies near the black hole ($\sim 100$ Hz)?

The inner MDAF model provides natural answers to these questions.
While the thin accretion disk is truncated by the ADAF at, perhaps,
$\sim 1000~r_{g}$, the ADAF itself is truncated at $R_1 \sim 100~
r_{g}$ by the MDAF, cutting off the ADAF turbulence above a few Hz.
The slope of this cutoff may represent the high frequency tail of the
turbulence spectrum near $R_1$.  The QPO is produced by the magnetic
flux tubes that stretch toward the black hole and rotate at roughly the
orbital $R_1$ frequency --- again a few Hz.

The power spectrum at each disk radius should be dominated by a rather
narrowly-peaked {\em local} spectrum \cite{MB02}.  We therefore can
approximate the {\em total} disk power spectrum by assuming the local
spectrum to be a delta function and simply plotting the variation of
the turbulence {\em strength} with radius against the variation of the
principal local (orbital) frequency with radius:
\begin{equation}
P[r(\nu)] = (2 \pi \, \Delta R \, \delta H) \, \rho \, V_{turb}(R)^2
\end{equation}
where $\Delta R \sim R$ is the annulus over which $\rho$ and $V_{turb}$
remain roughly constant, and $\delta H$ is the skin depth over which
the turbulent eddies can be seen by the observer (roughly the optical
depth). For an $\alpha$-disk \cite{SS73}, we find $P(r) \propto R^{1/2}
\propto \nu^{-1/3}$, or $dP/d{\nu} \propto \nu^{-4/3}$, in agreement with
GRS 1915+105 in the high state. However, in the low state, for a simple
evaporative ADAF model ($\dot{M} \propto R^{-1}$) \cite{Esin97}, we find
that $dP/d{\nu} \propto \nu^{-2/3}$, which is {\em not flat}.  In order
to obtain $dP/d{\nu} \propto \nu^{0}$, we need to assume a steeper rate
of evaporation of the thin disk into the ADAF:  $\dot{M} \propto R^{-2}$.
Detailed modeling of the power spectrum as disk turbulence at different
radii may, therefore, become an important diagnostic of conditions in
the optically thin portion of the accretion flow, ADAF and MDAF alike.

Figure 1 shows a schematic picture of GRS 1915+105 in the soft and hard
states, the corresponding power spectra, and energy inputs to and outputs
from the transition region.

\subsection{Relation to Presentations on XRBs at this Meeting}

\inlinecite{Fen04} have proposed a phenomenological model for jet
production in which the jet speed increases as the inner disk radius
decreases.  This model explains why strong jet outbursts are seen when
the disk transitions to the high/soft state and not when it transitions
to the low/hard state:  the jet speed {\em decreases} with time in the
latter case, resulting in no formation of a shock.

This model fits well with the MDAF scenario.  We identify the low/hard
state as one in which the cool accretion disk completely evaporates
before the ADAF transitions into an MDAF, {\it i.e.} for $R > R_1$.
In this case, near the central engine, the jet speed is simply the escape
speed from the transition radius, or
\begin{equation}
\label{eq:vjet}
V_{jet} \sim 2 G M / R_1 
\end{equation}
which gives a non-relativistic jet of $V_{jet} \sim 0.1~c$. Eventually,
as the accretion rate is increased, the cool disk begins to extend {\em
inside} $\sim 100~r_{g}$, and the ADAF changes from an accretion
flow in its own right to simply a corona above a dense cool disk.
The ADAF no longer extends inside the truncated cool disk; that region
is filled with the MDAF only, extending from the ADAF corona inward.
This begins the move toward the high state along the upper horizontal
branch in the intensity/hardness plane: the hard ADAF emission begins to
be suppressed, the thermal emission from the cool disk gains in strength,
the radius $R_1$ where the transition to MDAF occurs now follows the cool
disk truncation radius.  The jet velocity from equation (\ref{eq:vjet})
increases as the disk truncation radius decreases.  Eventually $R_1$
reaches all the way to the black hole horizon, and the MDAF is swallowed.
This turns off the jet, but not before its velocity reaches close to $c$
as $R_1 \rightarrow r_{H}$, the horizon radius.  It is this fast jet
that creates the shock and outburst that we observe.

\inlinecite{MMF04} also have interpreted the variability of XTE J1118+480
as coupling between the corona and the jet through a common reservoir
where large amounts of accretion power are stored.  In the MDAF model we
identify the transition region at $R_1$ as this reservoir.  Energy and
angular momentum input into this region comes from two sources:  the
accretion flow from outside and the magnetic coupling to the black hole
from inside $R_1$.  The output power is the jet production that occurs at
this radius.  It is important to note that the predicted temperature at
this transition region is of order a few $\times 10^9$ K, and it lies at
$\sim 100~r_{g}$ in the low state, but can move inward as the accretion
rate increases (see above).

\subsection{MDAFs and Low-Luminosity AGN}

Black hole accretion in active galactic nuclei (AGN) is expected to act similarly 
to that in XRB systems:  bright Seyfert and quasar objects are 
believed to be in a soft state while those AGN without strong 
optical line emission (low-luminosity AGN [LLAGN], FR I radio galaxies, 
Sgr A) are believed to be in a low/hard state.  While there is 
some timing data available on these latter objectes, a detailed 
comparison with the MDAF model is not possible at this time as a 
QPO-producing plateau state has not yet been identified.  Our 
discussion of MDAFs in AGN therefore will be more speculative.

LLAGN do indeed show bandwidth-limited noise, and the cutoff/break at high 
frequency sometimes is used as an indicator of black hole mass, with 
$\tau_{br} \approx 7.7 {\rm d} \, (M_{\bullet}/10^7 M_{\odot})$, where $\tau_{br}$ 
is the time scale, in days, where the break occurs in the AGN X-ray power 
fluctuation spectrum \inlinecite{Pap04}.  The MDAF model provides a physical reason why 
this {\it ad hoc} scaling of the break in different systems is a reasonable 
black hole mass indicator.  In the model the frequency of this break will 
be equal to, or slightly greater than, $\nu_A$ (equation \ref{eq:nuA}), so 
$\tau_{br} = 1 / \nu_A \approx 10 {\rm d} \, (M_{\bullet}/10^7 M_{\odot})$. 

AGN also display another property similar to that shown by X-ray binaries,  
and the MDAF model provides the same explanation there as well.  Jets produced 
by quasars and many Seyferts tend to be quite relativistic, even within only 
a parsec from the black hole core.  They therefore may be launched and 
accelerated rather close to the central black hole.  This suggestion is supported 
by semi-analytic jet acceleration models, which suggest a magnetic foot point 
only a few gravitational radii from the hole for 3C 345 \cite{VK04}.   However, jets 
produced by LLAGN and FR Is (and their counterparts, the BL Lacertae objects) 
are either less relativistic or show no motion at all.  A similar model for 
acceleration of the NGC 6251 jet yields an inner foot-point of $\sim 34 \, r_{g}$ 
for a $6 \times 10^8 \, M_{\odot}$ black hole \cite{VK04}.  Furthermore, M87 
shows significant collimation on scales of $60-200 \,r_{g}$ \cite{BJL02}, and
its jet speed at a distance of 0.16 pc from the core is only 0.1 c.  Yet, at kiloparsec 
distances, M87 shows {\em superluminal} motions up to 6 c.  Jets in AGN systems 
identified with the low/hard state appear to be launched with smaller velocities 
and at larger distances from the central black hole.  

It appears possible, then, that the jet-production region in LLAGN and FR I 
objects also may look like that in Figure 1, with the launch point lying many 
tens of gravitational radii from the black hole.  Only through continual, and 
persistent acceleration by the black hole over large vertical distances (many 
parsecs to kiloparsecs) do jets in low/hard state AGN achieve the relativistic 
speeds observed very far downstream of the accretion disk.

\section{Conclusions}

Interpretation of the photon and power spectra of black hole systems
like GRS 1915+105 leads to a new magnetically-dominated accretion flow
(MDAF) model for the low/hard state with three distinct disk regions:
\begin{enumerate}
\item{As in previous models, the outer region of the disk is a
geometrically thin, optically thick, and cool turbulent disk, driven by
the MRI.}
\item{Likewise, at intermediate radii ($\sim 100 - 1000~r_{g}$) there
is a {\em one}-temperature, advection-dominated, turbulent accretion flow
(ADAF) disk/corona that is geometrically thick, optically thin, and hot.
An evaporation rate into this corona that scales as $\dot{M} \propto
R^{-2}$ is more consistent with the power spectrum than other models.}
\item{The structure inside $\sim 100~r_{g}$ distinguishes this model
from others:  at the radius where cooling by relativistic electrons
becomes important, the ADAF transitions to an MDAF with $\alpha >>1$.
The inflow is extremely {\em in}efficient, non-turbulent and nearly
radial along strong magnetic field lines --- essentially an inward-facing
magnetosphere.  The narrow annulus where the flow transitions from
ADAF to MDAF is an ideal site for open field lines and the launching an
MHD-powered jet.}
\end{enumerate}
We identify the bandwidth-limited noise that appears in the low/hard state
as the ADAF's MRI turbulence viewed through the optically thin flow.
The MDAF model predicts the observed truncation of that noise at a few
Hz and the appearance of a strong QPO at the same place, as well as
the very low frequency QPOs at 0.01 - 0.1 Hz. Finally, the MDAF model
is consistent with the pheomenological models of \inlinecite{Fen04} and
\inlinecite{MMF04} and provides a physical connection between them and
black hole accretion theory.  In particular, extension of this model to
include an MDAF inside {\em all} truncated disks naturally predicts the
variation in jet speed with inner radius deduced by \inlinecite{Fen04}.

The model suggests a new interpretation of the power spectrum of
black hole candidates:  like the photon spectrum, each small range in
frequency $\Delta \nu$ is contributed by a given annulus $\Delta R$
in the accretion disk, with the central frequency corresponding to the
Keplerian frequency at that radius.  The spectral slopes are due not to
the physics of the turbulence itself but rather to variations in disk
{\em structure} with radius.  Only the cutoff at a few Hz is indicative
of the (high frequency end of the) local power spectrum there.

\acknowledgements

The author is especially grateful to T. Macarone and R. Fender for
organizing this conference on black hole accretion on all mass scales.
Emphasizing the similarities and scaling of black hole systems contributes
greatly to their overall understanding.  The author is supported, in part,
by a NASA Astrophysics Theory Program grant.  This research was performed
at the Jet Propulsion Laboratory, California Institute of Technology,
under contract to NASA.


\theendnotes

\end{article}
\end{document}